\documentclass[aps,prb,reprint,superscriptaddress,longbibliography]{revtex4-2}

\usepackage[T1]{fontenc}
\usepackage{lmodern}
\usepackage[svgnames]{xcolor}
\usepackage{lipsum}
\usepackage{graphicx}
\usepackage{cancel}
\usepackage{soul}

\definecolor{Dark}{gray}{0.2}
\definecolor{MedDark}{gray}{0.4}
\definecolor{Medium}{gray}{0.6}
\definecolor{Light}{gray}{0.8}
\definecolor{darkred}{rgb}{0.55, 0.0, 0.0}
\definecolor{darkslateblue}{rgb}{0.28, 0.24, 0.55}
\definecolor{royalblue(web)}{rgb}{0.25, 0.41, 0.88}
\usepackage{graphicx,float}
\usepackage{amsmath}
\usepackage{amssymb}
\usepackage[colorlinks]{hyperref}
\hypersetup{
    citecolor=darkred,
    linkcolor=blue,   
    urlcolor=blue}

\usepackage{bbm}

\def\be{\begin{equation}}
\def\ee{\end{equation}}
\def\beq{\begin{equation}}
\def\eeq{\end{equation}}
\def\bea{\begin{eqnarray}}
\def\eea{\end{eqnarray}}
\def\nbea{\begin{eqnarray*}}
\def\neea{\nonumber\end{eqnarray*}}
\def\bmat#1{\left(\begin{array}{#1}}
\def\emat{\end{array}\right)}
\def\bcase#1{\left\{\begin{array}{#1}}
\def\ecase{\end{array}\right.}
\def\bmini#1{\begin{minipage}{#1\textwidth}}
\def\emini{\end{minipage}}

\usepackage{wrapfig}
\graphicspath{{figures/}}
\usepackage{enumerate}

\usepackage{physics}

 

\begin{document}

\title{Understanding Anomalous Magnetothermal Transport via Disentangling Shear and Compression Phonons}
\date{\today}
\author{Haoting Xu}
\affiliation{Department of Physics, University of Toronto, 60 St. George St., Toronto, Ontario, Canada M5S 1A7}
\author{Antoine Matar}
\affiliation{Department of Physics, University of Toronto, 60 St. George St., Toronto, Ontario, Canada M5S 1A7}
\author{Hae-Young Kee}
\email[]{hy.kee@utoronto.ca}
\affiliation{Department of Physics, University of Toronto, 60 St. George St., Toronto, Ontario, Canada M5S 1A7}
\affiliation{Canadian Institute for Advanced Research, CIFAR Program in Quantum Materials, Toronto, Ontario, Canada M5G 1M1 }

\begin{abstract}
\noindent\textbf{\large Abstract} \par
\noindent Magnetothermal transport in various frustrated magnets exhibits striking field-dependent anomalies that deviate from conventional magnon or phonon transport.  To understand such anomalies, we derive an effective spin–phonon Hamiltonian in which phonons with different polarizations couple selectively to distinct spin operators in the strong spin–orbit coupling limit, and show that symmetry-constrained spin–lattice coupling naturally leads to mode-selective spin–phonon interactions.
As a result, compression and shear phonon modes contribute to heat current across different magnetic-field regimes. Using a Landauer transport framework combined with exact diagonalization of spin chains coupled to a phonon bath, we show that this mechanism produces a characteristic peak–dip–peak structure in the field dependence of heat current, providing a microscopic explanation for field-induced transport anomalies in spin-orbit-coupled Mott insulators.
\end{abstract}

\maketitle
\noindent\textbf{\large Introduction} \par
\noindent
Understanding thermal transport in insulating magnetic materials provides a powerful probe of low-energy excitations in quantum matter~\cite{thermal_transport_probing_quantum_material_review_Li_2020,thermal_Hall_review_2024,QSL_review_2020}. In conventional paramagnetic insulators, heat is primarily carried by phonons, and the thermal conductivity is largely insensitive to external magnetic fields~\cite{Slack_1973_thermal_review}. In magnetically ordered insulators, additional heat transport arises from magnons, typically enhancing the thermal conductivity at low temperatures~\cite{heat_transport_low_quantum_magnet_review_Hess_2019,large_magnon_heat_Cu_2017}. Under applied magnetic fields, magnon transport is suppressed due to the opening of a Zeeman gap, resulting in a monotonic decrease in magnon thermal conductivity~\cite{YIG_magnons_field_PRB_2015,magnon_mean_free_path_YIG_PRB_2014,YIG_large_field_PRB_2020,thermal_CrCl3_PRR_2020,li2025thermal_review_2}.

Recent experiments have revealed striking deviations from this conventional behavior. For example, in the strongly spin–orbit coupled magnet $\alpha$-RuCl$_3$, oscillatory longitudinal thermal conductivity has been observed under in-plane magnetic fields~\cite{oscillations_thermal_conductivity_2021,robustness_thermal_2021,RuCl3_origin_Takagi,RuCl3_sample,Oscillation_RuCl3_Young_June,RuCl3_stacking_disorder,RuCl3_ultraclean_single_crystal}. Such anomalies are often interpreted as signatures of exotic spinon Fermi surface~\cite{pseudoscalar_U1SL_RuCl3,aniso_RuCl3_KSL,RuCl3_Majorana_origin,RuCl3_top_phase_diagram_Li_Ern_Chern,generalized_Kitaev_model_Heqiu,fractionalized_ultrasound_PRB_2024}. However, the presence of a spin-liquid phase in these materials remains debated, motivating alternative explanations for these transport anomalies. Unconventional thermal transport has also been reported in various frustrated magnets and quantum spin liquid candidates~\cite{thermal_conductivity_spin_peierls_PRB1998,thermal_conductivity_spin_ladder_PRL2000,thermal_conductivity_spinon_PRB_2000,magnon_heat_LaCuO_PRL2003,thermal_conductivity_qsl_Nature_2009,heat_current_spin_liquid_Science_2010,thermal_Hall_neutral_spin_Science_2015,heat_qsl_kagome_volborthite_pnas2016,spin_thermal_Hall_Kogome_AFM_2018,thermall_hall_cuprate_Mott_insulator_NC2020,qsl_heat_1TTaS2_PRR_2020,spinon_transport_PbCuTe2O6_PRL2023,triangular_qsl_thermal_PRR2024,Ising-qsl_PRB2024,qsl_thermal_conductivity_PRB2025}.

\begin{figure}
    \centering
    \includegraphics[width=1.0\linewidth]{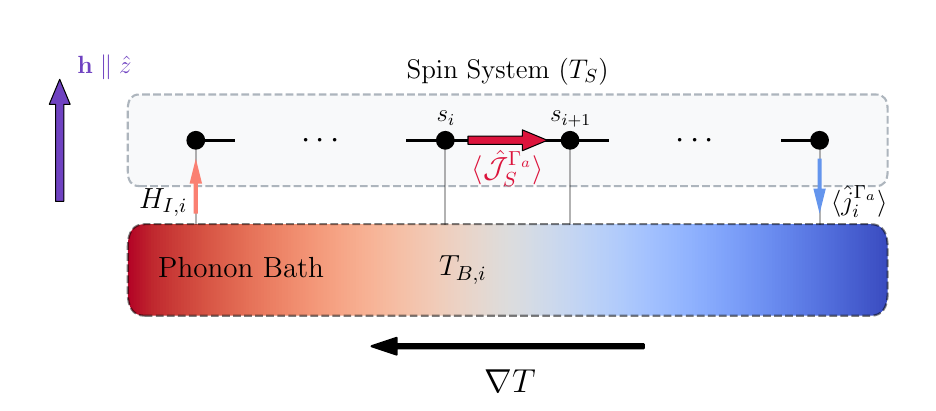}
    \caption{Schematic of the heat transport setup. A spin system (top) with an unperturbed temperature $T_S$ is coupled to a phonon bath (bottom) with a local temperature $T_{B,i}$ via the spin-phonon interaction $H_{I,i}$. An external magnetic field $\vb{h}\parallel \hat{z}$ is applied to the system, which induces the spin-phonon coupling. A spatial temperature gradient in the phonon bath creates a local temperature difference, driving a steady-state longitudinal heat current $\langle\hat{\mathcal{J} }_{S}^{\Gamma_a}\rangle$ at the central bond. The spin system absorbs energy at the hot end (red) and emits energy at the cold end (blue).  }
    \label{fig:spin-bath}
\end{figure}

Here we derive the effective spin–lattice coupling Hamiltonian with strong spin–orbit coupling (SOC) and show that crystal symmetry leads to a mode-selective spin–phonon interaction. Although magnetoelastic effects are widely recognized in thermal transport~\cite{spin_phonon_thermal_Walton_PRB_1970,magnon_phonon_Walton_PRB1977,thermal_conductivity_spin_chain_2003,giant_supression_phonon_heat_BiCu2PO6_Sci_reports_2016,review_spin_chain_heat_2021,MengxingYe_PHV_magnetic_insulators,resonant_phonon_scattering_Ni3TeO6_Heejun_PRB2022,phonon_dynamics_KSL_mengxing_ye,magnetoelastic_coupling_RuCl3_2021,heat_hybridization_Minhyea_Lee_PRR_2025}, the role of phonon polarization in driving energy currents remains largely unexplored. Because spin–lattice coupling is constrained by crystal symmetry, this raises the question of whether phonons with different polarizations couple selectively to distinct spin operators, leading to mode-dependent contributions to thermal transport across different field regimes.
While we are motivated by the anomalous thermal transport observed in $\alpha$-RuCl$_3$, the objective of the present work is not to quantitatively reproduce the experimental phenomenology of that material. Rather, we develop a general theoretical framework and demonstrate its implementation in one-dimensional spin systems, where the full spin excitation spectrum can be obtained with controlled numerical accuracy. Possible extensions of the formalism to higher-dimensional systems will be discussed in the Discussion section.

We show that longitudinal spin fluctuations generate compression-mode contributions to the heat current in the low-field frustrated phase, while transverse fluctuations contribute via shear modes below and above the transition, producing the anomalous heat current.
To illustrate this mechanism, we study several one-dimensional (1D) spin chains coupled to a phonon bath (Fig.~\ref{fig:spin-bath}) using a Landauer framework for nonequilibrium transport combined with exact diagonalization of the spin dynamics, revealing a nonmonotonic peak–dip–peak structure.

\vspace{0.5em}
\noindent\textbf{\large Results} \par
\noindent\textbf{Derivation of spin-phonon coupling} \par
\begin{figure}
    \centering
    \includegraphics[width=1.0\linewidth]{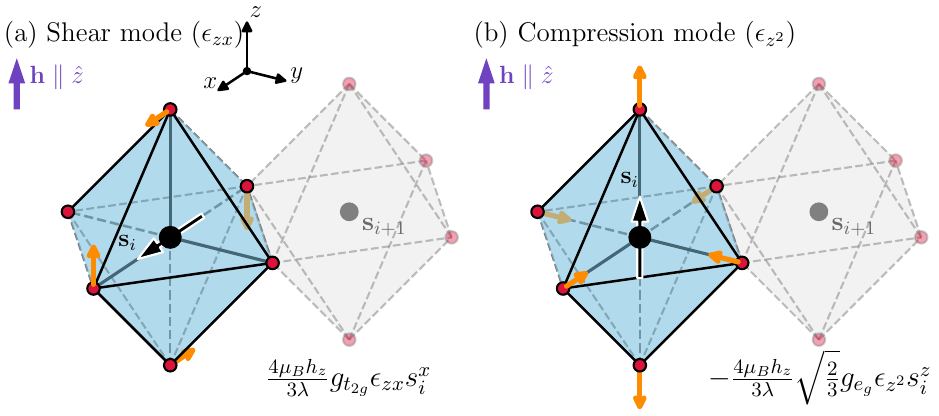}
    \caption{Schematic of the field-induced spin-lattice coupling in an edge-sharing octahedral structure. Under an external magnetic field $h_z$, distinct lattice distortions (orange arrows) selectively couple to different spin components of the central magnetic ion $\vb{s}_i$ (black arrows). (a) The shear mode ($\epsilon_{zx}$) couples to the transverse spin $s_i^x$. (b) The compression mode ($\epsilon_{z^2}$) couples to the field-aligned spin $s_i^z$. }
    \label{fig:phonon_modes}
\end{figure}

 \noindent
The linear coupling between spin and phonon operators in the presence of a magnetic field $\vb{h}$ in paramagnetic insulators was originally derived by Van Vleck \cite{VanVleck_1940}, Mattuck and Strandberg \cite{spin_phonon_Mattuck_1960}. In contrast to their approach, we focus on the case with large spin-orbit coupling \cite{Mott_insulator_Kitaev_Jackeli}. As an example, we consider $d^5$ or $d^7$ electron configuration surrounded by octahedral anions. The local ground state is the $J_{\rm eff}=1/2$ due to the spin-orbit coupling, resulting from total spin-1/2 or spin-3/2 and electronic $t_{2g}$ orbitals with $L_{\rm eff}=1$.\cite{Mott_insulator_Kitaev_Jackeli}
The lattice strain field couples to the electron degree of freedom by modulating of the crystal field potential. By symmetry analysis, the electron-strain coupling have the following form, 
\begin{equation}
\label{eq:electron-strain}
    H_{\rm el-strain}= \sum_{\Gamma_a} g_{\Gamma_a} \epsilon_{\Gamma_a} \mathcal{Q}_{\Gamma_a},
\end{equation}
where $\Gamma_a$ represents the $a$-component of the irreducible representation (irrep) $\Gamma$ of the lattice point group and its components. By symmetry, the coupling strength is independent of the component $a$ within a given irrep, $g_{\Gamma_a} \equiv g_\Gamma$. 
$\epsilon_{\Gamma_a}$ is the strain operator, and $\mathcal{Q}_{\Gamma_a}$ is the corresponding electron operator. Within the $t_{2g}$ orbitals arising from an octahedral crystal field, $\mathcal{Q}_{\Gamma_a}$ correspond to a quadrupole operator of the electron effective angular momentum $L_{\rm eff}=1$. 
The lattice strain couples to the $J_{\rm eff}=1/2$ pseudospin degrees of freedom (denoted by ${\vb{s}}$) through the interplay of SOC ($\lambda$) and an external magnetic field (${\bf h}$). Without loss of generality, we align the magnetic field along the $z$-direction ($h_z$). Treating the strain and magnetic field perturbatively, we derive the effective spin–lattice coupling Hamiltonian for a magnetic ion at site $i$ surrounded by octahedra cage as shown in Fig. \ref{fig:phonon_modes},
\begin{equation}
\label{eq:H_I,i}
    H_{I,i} = \frac{4\mu_B h_z}{3\lambda} \left[- \sqrt{\tfrac{2}{3}} g_{e_g} \epsilon_{z^2} s_i^z + g_{t_{2g}} \epsilon_{zx} s_i^x + g_{t_{2g}}\epsilon_{yz} s_i^y \right]. 
\end{equation}
Here, $\mu_B$ is the Bohr magneton, and $\Gamma_a = {z^2}$ represents the $e_g$ compression mode of the octahedral anions, while  $zx$ and $yz$ denote the $t_{2g}$ shear modes.
Detailed derivations are presented in the Supplemental Material (SM).
Eq.~\eqref{eq:H_I,i} reflects the crystal symmetry constraints on the spin-lattice coupling. As illustrated in Fig.~\ref{fig:phonon_modes}, for $\vb{h} \parallel \hat{z}$, the compression mode couples to the longitudinal pseudospin $s_i^z$, while shear modes couple to the transverse components $s_i^x$ and $s_i^y$. 
Although our analysis is based on an octahedral point group, it is applicable to crystal structures with other point groups. For $\alpha$-RuCl$_3$, first-principle calculations estimate $\tfrac{g_{t_{2g}}}{\lambda} \sim \tfrac{g_{e_g}}{\lambda} \sim \mathcal{O}(1)$.~\cite{magnetoelastic_coupling_RuCl3_2021}
As we demonstrate below, these distinct coupling forms lead to qualitatively different heat current signatures. 

Having established the local spin-lattice coupling, we derive the spin-phonon interaction by expressing the local strain tensor using second-quantized phonon operators. The lattice strain field operator can be written as 
\begin{equation}
    \epsilon_{\Gamma_a}(\vb{r})  = \sum_{\alpha,\vb{k}} \sqrt{\frac{\hbar}{2M \omega_{\alpha,\vb{k}}}} \phi^{\epsilon_{\Gamma_a}}_{\alpha,\vb{k}} (b_{\alpha,\vb{k}} + b_{\alpha,-\vb{k}}^\dagger ) e^{i \vb{k}\cdot \vb{r}_i},
\end{equation}
where $M$ is the ion mass, and $b_{\alpha,{\bf k}}$ and $b^\dagger_{\alpha,-{\bf k}}$ are annihilation and creation operators, respectively, with the phonon dispersion $\omega_{\alpha,{\bf k}}$ of the branch index $\alpha$ and the wavevector $\vb{k}$. Using the displacement $u_{\alpha,{\bf k}}$, $\phi^{\epsilon_{\Gamma_a}}_{\alpha,\vb{k}}  \equiv  \pdv{\epsilon_{\Gamma_a}}{u_{\alpha,\vb{k}}}$ denotes the transformation matrix relating the phonon eigenmodes to the strain components. 
In the second quantized form, the phonon Hamiltonian and the spin-phonon interaction can be written as,
\begin{equation}
\begin{aligned}
    H_{B} &= \sum_{\alpha \vb{k}} \hbar \omega_{\alpha \vb{k}} b_{\alpha \vb{k}}^\dagger b_{\alpha \vb{k}}, \\
    H_I &= \sum_{\Gamma_a} \sum_{i,\alpha \vb{k}} \gamma_{i,\alpha \vb{k}}^{\Gamma_a} (b_{\alpha,\vb{k}}+ b_{\alpha,-\vb{k}}^\dagger)s_i^{\mu}, \\
\end{aligned}
\label{eq:full_hamil}
\end{equation}
 where $\mu\in \{x,y,z\}$ denotes the spin component selected by the phonon mode $\Gamma_a$. 
The coupling $\gamma_{i,\alpha \vb{k}}^\mu$ takes the form
\begin{equation}
    \label{eq:spin-phonon-coupling-param}
    \gamma_{i,\alpha \vb{k}}^{\Gamma_a} = \frac{4\mu_B h_z g_{\Gamma}}{3\lambda}  \sqrt{\frac{\hbar}{2M \omega_{\alpha\vb{k} }}} \phi^{\epsilon_{\Gamma_a}}_{\alpha,\vb{k}} e^{i\vb{k}\cdot \vb{r}_i}, 
\end{equation}
where the numerical constants from Eq.~\eqref{eq:H_I,i} are absorbed into $g_{\Gamma}$.

\vspace{0.5em}
\noindent\textbf{Heat current via linear response} \par \noindent
The heat current at site $i$, which measures the energy transferring into the spin system, is obtained through energy continuity equation, 
\begin{equation}
    \sum_{i,\Gamma_a} \hat{j}_i^{\Gamma_a} = \frac{dH_{S}}{dt} = \frac{i}{\hbar}[H_S,H_I],
\end{equation}
where $\hat{j}_i^{\Gamma_a}$ is the energy current operator with specific strain mode $\Gamma_a$ and $H_S$ is the Hamiltonian of the system of interest. 
The full Hamiltonian is written as $H = H_S + H_B + H_I$, treating $H_I$ perturbatively. 
In our setup, the phonon environment is considered in the thermodynamic limit under a global temperature gradient. We treat the phonons as an effective thermal drive and handle the spin-phonon coupling perturbatively. The phonon modes coupled to the spin at site $i$ are described by an effective local thermal distribution with a site-dependent temperature $T_{B,i}$. This local-equilibrium description is introduced as an approximate framework for deriving the heat current. 
We further neglect spatial correlations between bath operators on different sites, as oscillating phase factors typically suppress non-local bosonic correlators. Consequently, we calculate the heat current on each site $\langle \hat{j}_i^{\Gamma_a}  \rangle$ for the nonequilibrium steady state using the following unperturbed density matrix, 
\begin{equation}
    \rho_{0,i} = \rho_{0,S}(T_S) \otimes \rho_{0,B} (T_{B,i}),
\end{equation}
where $\rho_{0,S}(T_S) = \sum_{m} Z_S^{-1} e^{-E_m/k_BT_S}\vert m \rangle \langle m \vert$ is the decoupled density matrix of the spin system at temperature $T_S$, $\rho_{0,B} (T_{B,i})$ denotes the phonon bath density matrix with local equilibrium temperature $T_{B,i}$, and $\vert m\rangle$ denotes the eigenstate of the spin system with eigenvalues of $E_m$. 

Using linear response theory, we calculate the steady-state heat current at local site $i$ with a small temperature difference between the spin system and the local equilibrium phonon bath. To linear order in the temperature difference, $\delta T_i = T_{B,i} - T_{S}$, the current is expressed via the Kubo formula,
\begin{equation}
    \langle \hat{j}_i^{\Gamma_a}  \rangle = \frac{\delta T_i}{k_BT_S^2} \int_0^{\infty} d t \left\langle \hat{j}_i^{\Gamma_a} (t) \hat{j}_i^{\Gamma_a} (0)\right \rangle_{\rho_{0,i}}.
\end{equation}
Evaluating the correlation functions in the eigenbasis of $H_S$, the current simplifies to 
\begin{equation}
    \label{eq:ji_calculate}
    \langle \hat{j}_i^{\Gamma_a}  \rangle =  \frac{ 2\pi\delta T_{i}}{k_B T_S^2} \int_0^\infty d \omega (\hbar \omega)^2 J_i^{\Gamma_a}(\omega) \mathcal{S}_{i,\mu}(\omega) n_{T_S}(\omega), 
\end{equation}
where $n_T(\omega)$ is the Bose-Einstein distribution with temperature $T$, $J_{i,a}^\Gamma(\omega)$ is the spectral density, and ${\cal S}_{i,\mu}(\omega)$ is the absorption spectral weight of the spin system defined below. 
This explicitly shows that if $\delta T_{i}>0$, 
energy flows into the system, satisfying the second law of thermodynamics. 
The detailed derivation can be found in the Methods. 

The spectral density of the spin-phonon coupling is defined as 
\begin{equation}
\label{eq:Jw_def}
    J_i^{\Gamma_a}(\omega) \equiv \sum_{\alpha,\vb{k}}\vert \gamma_{i,\alpha \vb{k}}^{\Gamma_a} \vert^2 \delta(\hbar \omega-\hbar \omega_{\alpha,\vb{k}}), 
\end{equation}
which encodes the spin-phonon coupling strength as well as the density of states of phonon. 
We evaluate the spectral density functions of the compressional and shear modes for a three-dimensional corner-sharing octahedral crystal structure, as shown in Fig. S1 of the SM. In the low-temperature regime, where acoustic modes dominate, the spectral density exhibits a characteristic cubic frequency dependence, $J_i^{\Gamma_a}(\omega) \propto \omega^3$. 
More complex phonon spectra and spin-lattice coupling strengths may shift the relative spectral weights of the compression and shear modes, but the qualitative scaling and the mode-selective mechanism remain robust.
The spin system absorption spectral weight, on the other hand, is defined as
\begin{equation}
\label{eq:emit_rate}
    \mathcal{S}_{i,\mu}(\omega) =  \sum_{m,n}  \frac{e^{- E_m/k_BT_{S}}}{Z_S} |\langle n|s_i^\mu|m\rangle|^2 \delta(\hbar\omega - E_{nm}),
\end{equation}
where $E_{nm} \equiv E_n - E_m$. 
This quantifies the thermally weighted transition probabilities across the spin energy levels resonant with the absorbed phonon energy from the phonon bath. 
To characterize bulk transport, we invoke steady-state local energy conservation to map the site currents $\langle\hat{j}_i^{\Gamma_a}\rangle$ onto the longitudinal bond current $\langle\hat{\mathcal{J}}_{S}^{\Gamma_a}\rangle$, which is evaluated at the chain center as shown in Fig. \ref{fig:spin-bath} by summing the heat current from the left side of the chain, i.e., $\sum_{i< N/2}\langle\hat{j}_i^{\Gamma_a}\rangle$.

Note that the heat current is fundamentally determined by the spectral overlap between two quantities, the phonon spectral density $J_i^{\Gamma_a}(\omega)$ and the spin system absorption spectral weight $\mathcal{S}_{i,\mu}(\omega)$. 
Eq.~\eqref{eq:ji_calculate} implies that 
distinct phonon strain modes $\Gamma_a$ couple exclusively to specific spin polarization operators $s_i^\mu$. In the low-field regimes hosting various magnetic phases, the low-energy excitation spectrum is governed by spin fluctuations both aligned and transverse with the external field. 
Consequently, both the compression and shear modes contribute to the thermal transport, unless the longitudinal fluctuation is completely suppressed by a magnetic order. 
As the system enters the fully polarized phase at high magnetic fields, the longitudinal spin fluctuations are frozen out. The surviving low-energy channels are spin flips. Thus, the transverse spectral weight takes over, causing the shear modes to emerge as the exclusive drivers of the heat current just above the phase transition. At higher field, when the Zeeman gap exceeds the thermal energy scale, the magnetothermal transport is suppressed. 
This nonmonotonic field-dependence persists when $k_B T \lesssim J$, and is suppressed when $k_B T \sim J$, at which point the system enters the paramagnetic regime. 

\vspace{0.5em}
\noindent\textbf{Application to 1D spin chains} \par \noindent
To demonstrate the proposed mechanism, we apply our transport framework to several one-dimensional (1D) spin chains ($H_S$). 
Given that SOC is explicitly incorporated in the spin-lattice coupling, we naturally focus on anisotropic spin models beyond the conventional isotropic Heisenberg model. Below, we present the heat current results for three representative cases -- antiferromagnetic (AFM) XXZ model, Kitaev-Heisenberg model, and the ferromagnetic (FM) XXZ model.

We first consider an extended AFM XXZ chain, which captures the SOC-induced Ising anisotropy~\cite{Churchill_2024_prl}. 
The spin dynamics are governed by the Hamiltonian
\begin{equation}
    H_{S} = \sum_{i=1}^N \big\{ J_i (s_i^x s_{i+1}^x + s_i^y s_{i+1}^y + \varepsilon s_i^z s_{i+1}^z ) - h_z s_i^z \big\},
\end{equation}
where $J_i = J[1+(-1)^i \delta]$, $J=1$ sets the energy scale of spin exchange coupling, $\delta$ parameterizes the bond alternation, and $\varepsilon$ dictates the Ising anisotropy. This 1D model exhibits a rich phase diagram driven by the external magnetic field $h_z$~\cite{phase_diagram_XXZ_chain,XXZ_alter_PRB_2024}. In the regime of strong anisotropy and weak dimerization, the system undergoes two successive quantum phase transitions. For low fields ($h_z < h_{c1}$), the ground state is a gapped dimerized phase dominated by nearest-neighbor singlets. Intermediate fields ($h_{c1} < h_z < h_{c2}$) stabilize a gapless Luttinger liquid (LL) phase, which ultimately gives way to a fully polarized state when $h_z > h_{c2}$. To quantitatively investigate the thermal transport, we adopt the specific parameters $\varepsilon = 3.0$ and $\delta = 0.2$. Density matrix renormalization group (DMRG) calculations pinpoint the critical fields at $h_{c1} = 1.0J $ and $h_{c2} = 3.0J $. 

\begin{figure}
    \centering
    \includegraphics[width=1.0\linewidth]{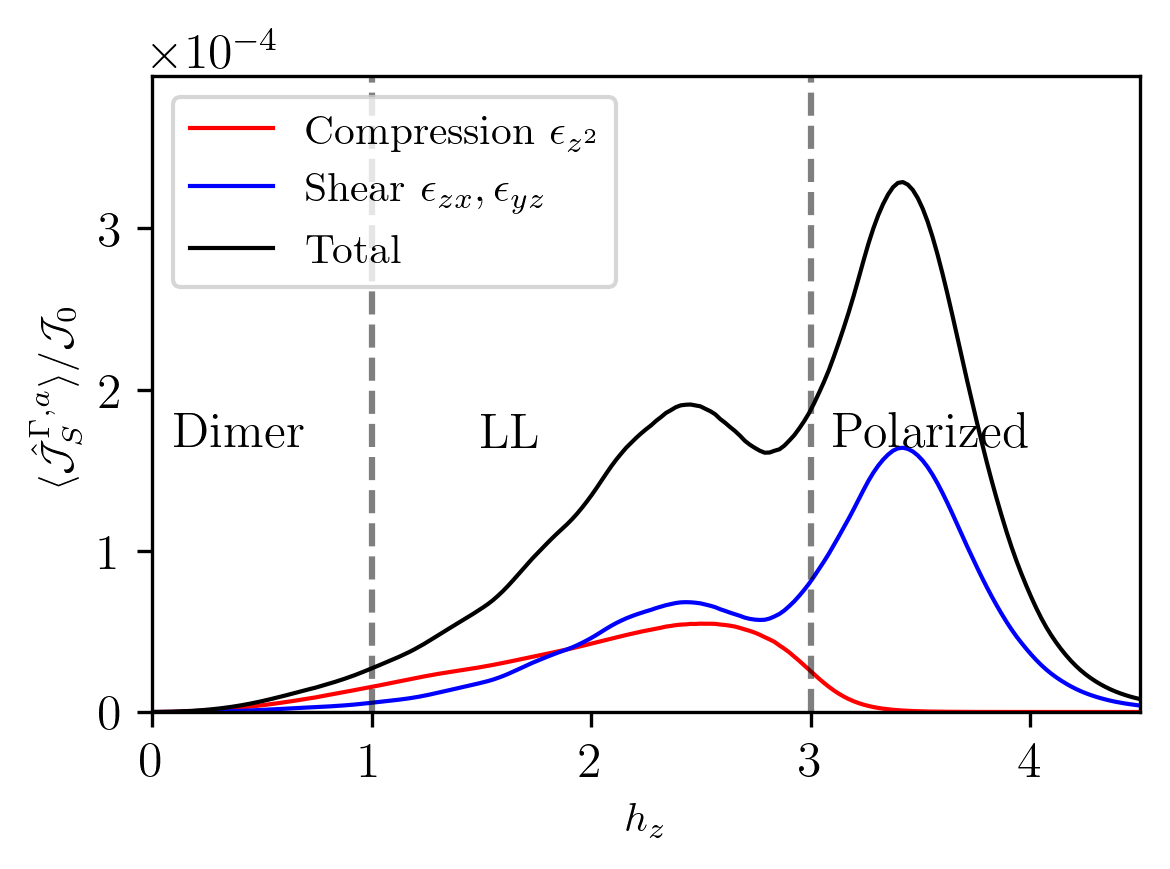}
    \caption{Field dependence of the heat current in the 1D XXZ chain in unit of the heat current scale $\mathcal{J}_0$ (see the definition and estimation in the SM). The total heat current (black curve) is decomposed into contributions from the compression strain mode ($\epsilon_{z^2}$, red) and the two shear modes ($\epsilon_{zx}, \epsilon_{yz}$, blue). The vertical gray dashed lines indicate the critical fields ($h_{c1}, h_{c2}$) separating the dimer, Luttinger liquid (LL), and polarized phases. Notably, whereas the LL phase involves contributions from all strain channels, the compression mode vanishes in the polarized regime, giving rise to the characteristic non-monotonic field dependence. }
    \label{fig:heat_current_XXZ}
\end{figure}

The spin absorption spectral weight, defined in Eq.~(\ref{eq:emit_rate}), is computed using exact diagonalization (ED) on an $N=20$ cluster. Crucially, the phase boundaries extracted from our ED spectra perfectly match the DMRG results, confirming that an $N=20$ system is sufficient to accurately capture the thermodynamic limit behavior. Finally, we evaluate the anomalous heat current mediated by distinct phonon symmetries (shear vs.\ compression modes), as depicted in Fig.~\ref{fig:heat_current_XXZ}. 
The average temperature of phonon bath and the system is set to $k_BT_S = 0.1J$, with a small thermal difference $\delta T = 0.05 T_S$ in the phonon bath to ensure the validity of the linear response regime. 

In Fig.~\ref{fig:heat_current_XXZ}, we illustrate the evolution of the anomalous heat current with the magnetic field $h_z$, highlighting the interplay between distinct phonon symmetries and the underlying low-field magnetic phases. Following Eq.~(\ref{eq:Jw_def}), the bath spectral density inherently scales as $h_z^2$ due to the spin-phonon coupling, ensuring that this anomalous thermal contribution strictly vanishes at zero field. In the low-field dimerized phase ($h_z < h_{c1}$), thermal transport is dominated by the compression mode coupling to $s_i^z$. Because the $s_i^z$ operator readily breaks the nearest-neighbor singlets, this channel yields a heat current that closely tracks the $h_z^2$ scaling of the bath. Conversely, the shear modes (coupling to $s_i^x$ and $s_i^y$) contribute negligibly, as the required transverse spin flips cost higher energy. Due to the additional scaling with $h_z^2$, both the modes contribute little current as compared to higher field regions.

In the gapless Luttinger liquid phase $(h_{c1}<h_z<h_{c2})$, the excitations become collective spin bosons, and both compression and shear modes contribute to the heat current through longitudinal and transverse spin fluctuations, respectively. The phase transition from the dimerized phase to the Luttinger liquid leaves no signature in the thermal current, because the change in the spin absorption spectrum is multiplied by the small $h_z^2$ prefactor at low magnetic fields and does not show up in the thermal current.
In the polarized phase $(h>h_{c2})$, the compression contribution vanishes while the shear mode develops a peak. Because the ground state is an eigenstate of $s_i^z$, longitudinal compression modes cannot generate magnetic excitations, whereas transverse shear mode flip polarized spins and produce a heat-current peak just above $h_{c2}$. At larger fields, the growing Zeeman gap suppresses magnetic excitations, leading to an exponential decay of the heat current. 

This behavior is generic when both longitudinal and transverse fluctuations are significant in the low-field phase. As demonstrated below for two additional 1D spin chains, the compression-mode contribution remains large in the Kitaev-Heisenberg chain, preserving the anomalous peak–dip–peak structure. By contrast, in the strongly ordered state of the FM XXZ model, the compression mode contribution is strongly suppressed, producing a large heat current just above the transition, confirming the key role of longitudinal fluctuations in promoting the anomalous heat current.

\begin{figure}
    \centering
    \includegraphics[width=1.0\linewidth]{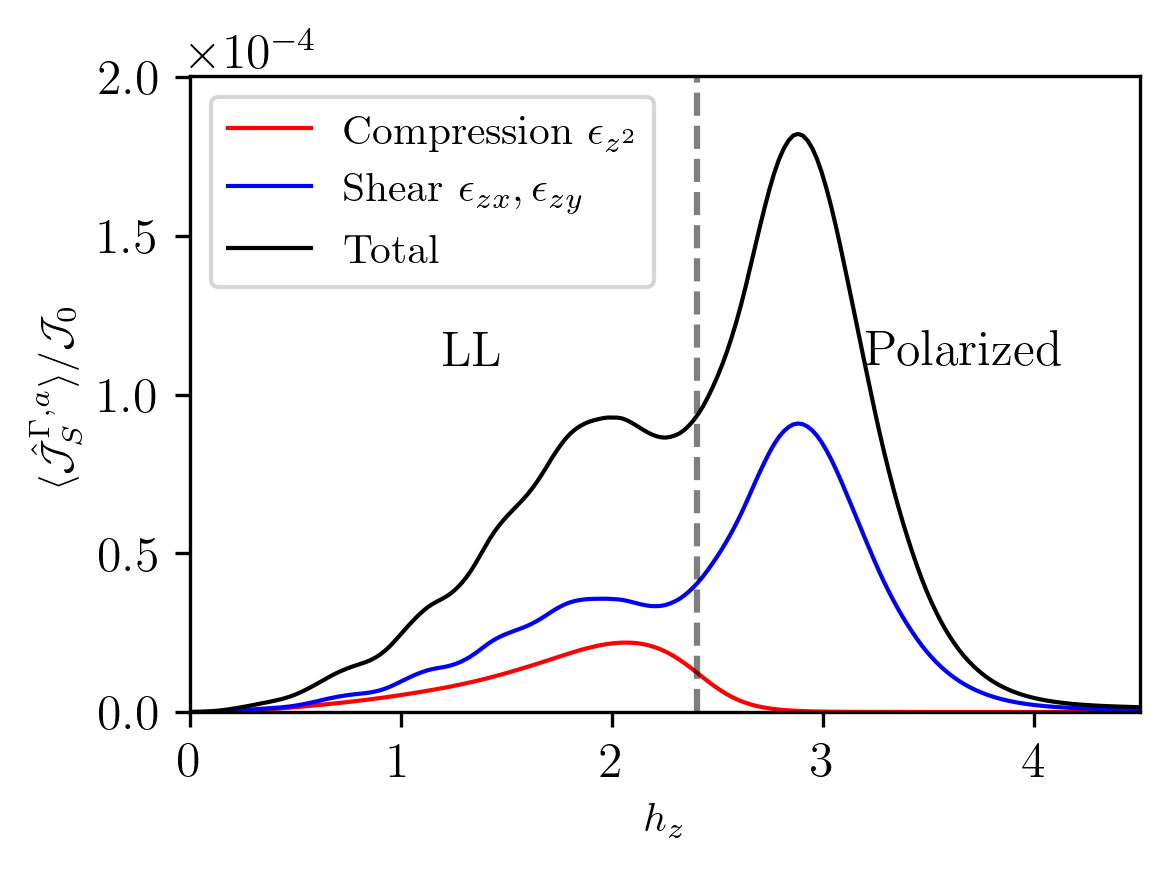}
    \caption{Field dependence of the heat current for the 1D Kitaev-Heisenberg chain. The total heat current (black) is decomposed into contributions from the compression strain mode ($\epsilon_{z^2}$, red) and the shear strain modes ($\epsilon_{zx}$, $\epsilon_{yz}$, blue). The vertical gray dashed line indicates the critical field separating the Luttinger liquid (LL) and the polarized phases. While the LL phase involves contributions from all strain channels, the compression-mode contribution vanishes in the polarized field regime, giving rise to the characteristic nonmonotonic field dependence. }
    \label{fig:KJ_chain_current}
\end{figure}
In order to demonstrate the generality of the heat current behavior, we investigate the Kitaev-Heisenberg spin chain as another example that can give rise to the anomalous peak-dip-peak structure. The Kitaev and Heisenberg spin interactions are present in several quasi-1D magnetic insulators, like CoNb$_2$O$_6$~\cite{Churchill_2024_prl}. The spin dynamics are governed by the Hamiltonian, 
\begin{equation}
H_{S} = \sum_{\langle i,j\rangle} \left( K s_i^\gamma s_j^\gamma +J \vb{s}_i \cdot \vb{s}_j \right) - \sum_{i=1}^N h_z s_i^z.
\end{equation}
Here, $\gamma \in \{x,y\}$ denotes the bond index of the Kitaev interaction, characterized by alternating nearest neighbor $x$-bonds and $y$-bonds. 
Building on the comprehensive phase diagram of the extended $K-J-\Gamma$ chain established by Yang \textit{et al.} \cite{comprehensive_KJGamma_chain_2020_Affleck}, we focus on the AFM interaction with $K=1.0$ and $J=1.0$. At zero field, the system realizes a gapless Luttinger liquid (LL) phase. Driven by a longitudinal magnetic field $h_z$, the LL undergoes a quantum phase transition into a fully polarized state. Employing DMRG simulations, we pinpoint this quantum critical point at $h_c = 2.4$. Exact diagonalization on a finite cluster of $N=20$ sites perfectly reproduces this critical field, demonstrating that our ED approach effectively captures the thermodynamic-limit physics. 

Following the setup detailed in Methods, the magnetic-field dependence of the magnetothermal current for the Kitaev-Heisenberg chain is presented in Fig.~\ref{fig:KJ_chain_current}. 
In the low-field gapless Luttinger liquid (LL) phase, both the compression and shear modes contribute to the thermal transport via longitudinal and transverse spin fluctuations. Notably, both modes exhibit a peak when the magnetization reaches half of its saturation value, corresponding to a maximum in the density of states for low-energy spin excitations. In the polarized phase ($h > h_c$), a distinct peak emerges for the shear mode while the compression-mode contribution vanishes, governed by the same selection mechanism in the AFM XXZ chain.

To explicitly confirm that low-field longitudinal spin fluctuations are essential for the anomalous peak-dip-peak structure, we examine the 1D FM XXZ chain as a contrasting example. The spin dynamics of the 1D FM XXZ chain under a transverse magnetic field are governed by the Hamiltonian
\begin{equation}
    H_{S} = \sum_{i=1}^N \left \{ -J (s_i^x s_{i+1}^x + s_i^y s_{i+1}^y + \varepsilon s_i^z s_{i+1}^z ) -h_x s_i^x  \right \},
\end{equation}
where $J = 1\ \mathrm{meV}$ sets the ferromagnetic exchange energy scale and $\varepsilon$ dictates the Ising anisotropy. Because the external field is applied along the $x$-axis, the symmetry-constrained spin-phonon coupling undergoes a corresponding spatial rotation. Consequently, the relevant compression mode is denoted as $\epsilon_{x^2}$ and couples to the longitudinal spin component $s_i^x$, whereas the shear modes $\epsilon_{xy}$ and $\epsilon_{zx}$ couple to the transverse components $s_i^y$ and $s_i^z$. At zero field, the ground state exhibits a ferromagnetic order along the $z$-axis. The large Ising anisotropy establishes a strong easy axis, such that increasing the transverse magnetic field drives the system through a single quantum phase transition. Setting $\varepsilon=2$ for our numerical simulations, we determine the quantum critical point to be $h_c^\prime = 0.52 J$ using DMRG and ED. For $h>h_c^\prime$, it becomes fully polarized along the applied field direction.

\begin{figure}
    \centering
    \includegraphics[width=1.0\linewidth]{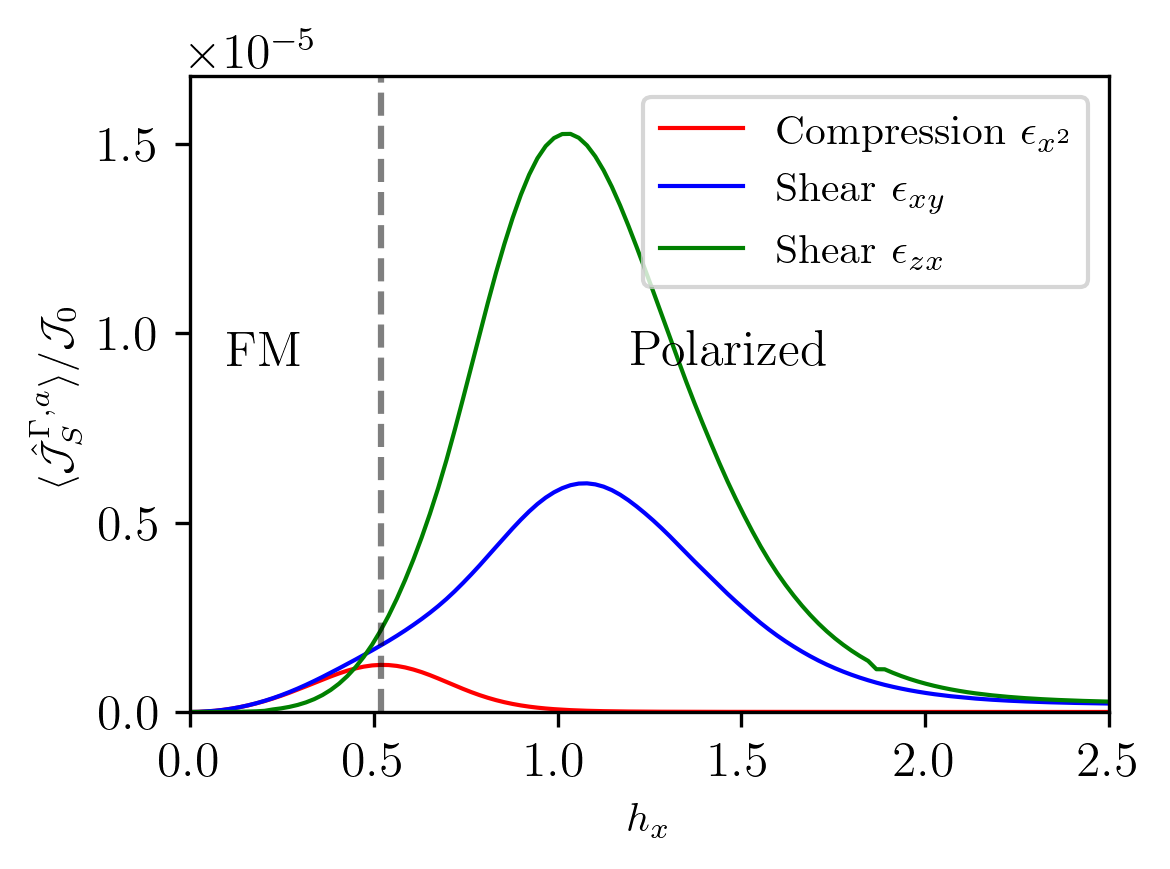}
    \caption{Field dependence of the heat current for the 1D ferromagnetic XXZ chain. The heat current is resolved into contributions from the compression ($\epsilon_{x^2}$, red) and shear ($\epsilon_{xy}$, blue; $\epsilon_{zx}$, green) strain modes. The vertical gray dashed line marks the critical point separating the ferromagnetic and polarized phases. Within the low-field ferromagnetic phase, the compression mode contributes little heat current as the longitudinal spin fluctuations are suppressed due to the ordered state. The total current does not show the anomalous oscillating structure.}
    \label{fig:SM-FM-XXZ}
\end{figure}

Fig.~\ref{fig:SM-FM-XXZ} illustrates the magnetic-field dependence of the heat current. Unlike the frustrated regimes, the low-field phase here is a strongly ordered ferromagnetic state, which strongly suppresses the longitudinal spin fluctuations ($s_i^x$). As a result, the compression mode (coupled to $s_i^x$) contributes little to the heat current. In contrast, the shear modes remain active due to transverse fluctuations ($s_i^y, s_i^z$) sustained by the XY exchange. Upon entering the fully polarized phase ($h_x > h_c^\prime$), the compression channel completely vanishes, and the shear mode produces a single pronounced peak.
Because the low-field phase lacks sufficient compression-mode contributions, the signature peak-dip-peak structure field dependence is absent. 
This result confirms the importance of significant longitudinal fluctuations in the low-field phase for generating the anomalous peak-dip-peak heat current.

\vspace{0.5em}
\noindent\textbf{\large Discussions} \par \noindent
We develop a microscopic theory of heat transport in which spins are coupled to a phonon bath under a temperature gradient with an applied magnetic field, uncovering a fundamental phonon polarization-selection mechanism.  In the strong spin-orbit-coupled limit, we show that distinct acoustic phonon modes couple selectively to specific spin components. 
Using the Landauer formalism and exact diagonalization, we find that while both compression and shear modes contribute to the heat transfer in weakly-ordered or frustrated phases in low-field regime, the high-field polarized phase exhibits a significant suppression of the longitudinal channel,
leaving the transverse shear modes to dominate the heat current. 
This mode-selective mechanism, enhanced by the combined contributions of the compression and shear modes, gives rise to the nonmonotonic peak-dip-peak field dependence of the heat current.

Our framework offers a new perspective on the field dependence of thermal conductivity in spin–orbit-coupled magnets. We show that nonmonotonic field responses can arise from the interplay between the mode-selective spin fluctuations and the underlying magnetic phases, providing an alternative explanation for anomalous transport.
The symmetry constraints further imply highly anisotropic spin–phonon interactions, suggesting angle-dependent thermal conductivity and extensions to higher-dimensional spin models as promising directions for future work.

Extending the same analysis to higher dimensional spin systems such as $\alpha$-RuCl$_3$, a layered two-dimensional honeycomb systems with substantial computational complexity, would require access to the full spin excitation spectrum in significantly larger Hilbert spaces. Despite these technical challenges, we believe that the present theoretical framework provides a useful foundation for the study of anomalous thermal transport in higher-dimensional systems owing to its general and broadly applicable formulation.  For example, multiple anomalous features observed in the longitudinal magnetothermal transport of $\alpha$-RuCl$_3$ under magnetic field have been discussed in terms of successive magnetic transitions induced by distinct magnetic phases or stacking-related effects~\cite{RuCl3_origin_Takagi,Cen2025PRB}.
Since we already incorporate the three-dimensional phonon spectrum in the present framework, if multiple magnetic transitions are present in spin systems, the longitudinal magnetothermal transport is expected to reflect these transitions through multiple anomalous peaks, similar to those reported in $\alpha$-RuCl$_3$. We anticipate that its application to such systems will be an important direction for future studies. 

\vspace{0.5em}
\noindent\textbf{\large Methods} \par
\noindent\textbf{Linear response theory for magnetothermal transport} \par \noindent
To evaluate the mangetothermal transport using linear response theory, we treat the composite spin-phonon system with the full Hamiltonian $H = H_S + H_B + H_I$. The unperturbed spin system is formally expressed as $H_S = \sum_m E_m \vert m \rangle \langle m\vert$, with $E_m$ and $\vert m\rangle$ being its eigenenergies and eigenstates. We rewrite the interacting Hamiltonian as 
\begin{equation}
    H_I = \sum_{\mu} \sum_{i,\alpha \vb{k}} \gamma_{i,\alpha \vb{k}}^{\Gamma_a} (b_{\alpha,\vb{k}}+ b_{\alpha,-\vb{k}}^\dagger)s_i^{\mu} \equiv \sum_{i,\mu} s_i^{\mu} B_i^{\mu},  
\end{equation}
where $B_i^\mu$ denotes the bosonic bath operator coupled to the local spin component $s_i^\mu$ via the specific strain mode $\Gamma_a$. 

The heat current at site $i$, which quantifies the energy transfer rate into the spin system, obeys the energy continuity equation $\sum_{i, \Gamma_a} \hat{j}_i^{\Gamma_a} = (i/\hbar)[H_S,H_I]$.
The current operator is defined as $\hat{j}_i^{\Gamma_a} = \dot{s}_i^\mu B_i^\mu$, where $\dot{s}_i^{\mu}$ denotes the time derivative of the spin operator in the interaction picture. 
To evaluate the steady-state expectation value of this heat current, we utilize the unperturbed density matrix $\rho_{0,i} = \rho_{0,S}(T_S) \otimes \rho_{0,B} (T_{B,i})$, where $\rho_{0,S}(T_S) = \sum_{m} Z_S^{-1} e^{-E_m/k_B T_S}\vert m \rangle \langle m \vert$ is the equilibrium density matrix of the unperturbed spin system at the global temperature $T_S$, and $\rho_{0,B} (T_{B,i})$ denotes the local phonon bath density matrix at temperature $T_{B,i}$. 

Assuming a small local temperature deviation $\delta T_i = T_{B,i} - T_S$ between the local phonon bath and the global spin system, the steady-state current to the linear order in $\delta T_i$ is obtained using the Kubo formula,
\begin{equation}
\begin{aligned}
    &\left\langle\hat{j}_i^{\Gamma_a}\right\rangle=\frac{\delta T_i}{k_BT_S^2 } \int_0^{\infty} d t\left\langle\hat{j}_i^{\Gamma_a}(t) \hat{j}_i^{\Gamma_a}(0)\right\rangle_{\rho_{0,i}}\\
&=  \frac{\delta T_i}{k_B\left(T_S\right)^2} \int_0^{\infty} d t\left\langle \dot{s}_i^{\mu}(t) \dot{s}_i^{\mu}(0) \right\rangle_{\rho_{0,S}} \left\langle B_i^{\mu}(t) B_i^{\mu}(0)\right\rangle_{\rho_{0,B}}. 
\end{aligned}
\end{equation}
The time evolution of the operators is governed by the unperturbed Hamiltonian in the interaction picture, such that $\dot{s}_i(t) =  e^{iH_S t} \dot{s}_i(0) e^{-i H_S t}$ and $B_i(t) =e^{iH_B t} B_i(0) e^{-i H_B t}$. Evaluating these correlation functions yields
\begin{equation}
\begin{aligned}
    \left\langle \dot{s}_i^\mu(t) \dot{s}_i^{\mu}(0) \right\rangle_{\rho_{0,S}} =&\sum_{m^\prime,m}Z_S^{-1} e^{-E_m/T_S} e^{i(E_{m}-E_{m^\prime})t} E_{m^\prime m}^2 \\ 
    &\times \vert \langle m \vert s_i^\mu \vert m^\prime \rangle \vert ^2,\\
        \langle B_i^{\mu}(t) B_i^{\mu}(0)\rangle_{\rho_{0,B}} =& \sum_{\alpha,\vb{k}} \vert \gamma_{i,\alpha \vb{k}}^{\Gamma_a}\vert^2 e^{-i\omega_{\alpha,\vb{k}}t} \left( n_{T_{B,i}} (\omega_{\alpha,\vb{k}}) +1   \right)\\& + \vert \gamma_{i,\alpha\vb{k}}^{\Gamma_a}\vert^2 e^{i\omega_{\alpha,\vb{k}}t} n_{T_{B,i}} (\omega_{\alpha,\vb{k}}).\\
\end{aligned}
\end{equation}
where $n_{T_{B,i}}(\omega)$ is the Bose-Einstein distribution. Substituting these expressions back into the Kubo formula and performing the time integration, the heat current simplifies to the results given in Eq.~\eqref{eq:ji_calculate}. 

\vspace{0.5em}
\noindent\textbf{Exact diagonalization and DMRG calculations} \par \noindent
To establish the ground-state phase diagram, we employ density matrix renormalization group (DMRG) calculations on a system of $N=100$ spins, precisely pinpointing the quantum critical fields for the given spin model. The phase boundaries extracted from the ground-state energy spectra of exact diagonalization (ED) on a finite cluster of size $N=20$ are in agreement with these DMRG results. This verifies that the $N=20$ cluster is sufficient to accurately capture the thermodynamic limit behavior across all field-induced phases. 
With the finite-size scaling validated, the spin dynamics and energy levels required for the transport evaluation are computed numerically using ED on the $N=20$ cluster. Specifically, we calculate the spin absorption spectral weight $\mathcal{S}_{i,\mu}(\omega)$ defined in Eq.~\eqref{eq:emit_rate}. To optimize the computational efficiency, the eigenspectrum is truncated to retain only states with energies up to $10 k_B T_S$. 
The spatial temperature gradient in the phonon bath is established by setting a local temperature difference $\delta T_i = \delta T (\frac{2(i-1)}{N-1}-1)$. To rigorously ensure the validity of the linear response regime, the unperturbed global temperature of the system is set to $k_BT_S = 0.1J$, and the maximum thermal difference across the bath is kept small at $\delta T = 0.05 T_S$. 

The resulting local site current $\langle\hat{j}_i^{\Gamma_a}\rangle$ is governed by the spectral overlap between the phonon bath spectral density $J_i^{\Gamma_a}(\omega)$ and the computed spin absorption spectral weight $\mathcal{S}_{i,\mu}(\omega)$. Finally, to characterize the bulk transport properties, we map these local site currents onto the macroscopic longitudinal bond current $\langle\hat{\mathcal{J}}_{S}^{\Gamma_a}\rangle$. This is evaluated at the center of the one-dimensional chain by summing the local heat current contributions from the left half of the system, taking the form $\sum_{i< N/2}\langle\hat{j}_i^{\Gamma_a}\rangle$.

\vspace{0.5em}
\noindent\textbf{\large Data Availability} \par \noindent
The numerical data supporting the findings of this study are openly available at the following URL: \url{https://doi.org/10.5281/zenodo.19499326}. 

\vspace{0.5em}
\noindent\textbf{\large Acknowledgements} \par \noindent
This work is supported by the NSERC Discovery Grant No. 2022-04601 and NSERC CREATE program No. 575280-2023. H. Y. K. acknowledges support from the Canada Research Chairs Program No. CRC-2019-00147.
This research was enabled in part by support provided by Calcul Québec and the Digital Research Alliance of Canada.

\vspace{0.5em}
\par\noindent\textbf{\large Author Contributions} \par \noindent
H.X. carried out the derivation of the effective spin–phonon coupling and the linear response theory, and performed the numerical calculations of the heat current. A.M. calculated the phonon spectrum and the spectral density. H.-Y.K. conceived the project and supervised the research. All authors discussed the results and contributed to writing the manuscript.

\vspace{0.5em}
\noindent\textbf{\large Competing Interests} \par \noindent
Author H.-Y.K. is an Associate Editor of npj Quantum Materials. H.-Y.K. was not involved in the journal’s review of, or decisions related to, this manuscript.

\bibliographystyle{naturemag}
\bibliography{cites}

@article{oscillations_thermal_conductivity_2021,
  title={Oscillations of the thermal conductivity in the spin-liquid state of {$\alpha$-RuCl$_3$}},
  author={Czajka, Peter and Gao, Tong and Hirschberger, Max and Lampen-Kelley, Paula and Banerjee, Arnab and Yan, Jiaqiang and Mandrus, David G and Nagler, Stephen E and Ong, NP},
  journal={Nature Physics},
  volume={17},
  number={8},
  pages={915--919},
  year={2021},
  publisher={Nature Publishing Group UK London},
  doi={https://doi.org/10.1038/s41567-021-01243-x},
  url={https://doi.org/10.1038/s41567-021-01243-x}
}

@article{robustness_thermal_2021,
  title={Robustness of the thermal Hall effect close to half-quantization in {$\alpha$-RuCl$_3$}},
  author={Bruin, JAN and Claus, RR and Matsumoto, Y and Kurita, N and Tanaka, H and Takagi, H},
  journal={Nature Physics},
  volume={18},
  number={4},
  pages={401--405},
  year={2022},
  publisher={Nature Publishing Group UK London},
  doi={https://doi.org/10.1038/s41567-021-01501-y},
  url={https://doi.org/10.1038/s41567-021-01501-y}
}

@article{Oscillation_RuCl3_Young_June,
  title = {Oscillations in the magnetothermal conductivity of {$\ensuremath{\alpha}\ensuremath{-}{\mathrm{RuCl}}_{3}$}: Evidence of transition anomalies},
  author = {Lefran\ifmmode \mbox{\c{c}}\else \c{c}\fi{}ois, \'Etienne and Baglo, Jordan and Barth\'elemy, Q. and Kim, S. and Kim, Young-June and Taillefer, Louis},
  journal = {Phys. Rev. B},
  volume = {107},
  issue = {6},
  pages = {064408},
  numpages = {7},
  year = {2023},
  month = {Feb},
  publisher = {American Physical Society},
  doi = {10.1103/PhysRevB.107.064408},
  url = {https://link.aps.org/doi/10.1103/PhysRevB.107.064408}
}

@article{aniso_RuCl3_KSL,
  title = {Anisotropic Thermal Conductivity Oscillations in Relation to the Putative {Kitaev} Spin Liquid Phase of {$\alpha$-RuCl$_3$}},
  author = {Zhang, Heda and Miao, Hu and Ward, Thomas Z. and Mandrus, David G. and Nagler, Stephen E. and McGuire, Michael A. and Yan, Jiaqiang},
  journal = {Phys. Rev. Lett.},
  volume = {133},
  issue = {20},
  pages = {206603},
  numpages = {8},
  year = {2024},
  month = {Nov},
  publisher = {American Physical Society},
  doi = {10.1103/PhysRevLett.133.206603},
  url = {https://link.aps.org/doi/10.1103/PhysRevLett.133.206603}
}

@article{RuCl3_sample,
  title = {Sample-dependent and sample-independent thermal transport properties of {$\alpha$-RuCl$_3$}},
  author = {Zhang, Heda and May, Andrew F. and Miao, Hu and Sales, Brian C. and Mandrus, David G. and Nagler, Stephen E. and McGuire, Michael A. and Yan, Jiaqiang},
  journal = {Phys. Rev. Mater.},
  volume = {7},
  issue = {11},
  pages = {114403},
  numpages = {6},
  year = {2023},
  month = {Nov},
  publisher = {American Physical Society},
  doi = {10.1103/PhysRevMaterials.7.114403},
  url = {https://link.aps.org/doi/10.1103/PhysRevMaterials.7.114403}
}

@article{RuCl3_Majorana_origin,
author = {Kumpei Imamura  and Shota Suetsugu  and Yuta Mizukami  and Yusei Yoshida  and Kenichiro Hashimoto  and Kenichi Ohtsuka  and Yuichi Kasahara  and Nobuyuki Kurita  and Hidekazu Tanaka  and Pureum Noh  and Joji Nasu  and Eun-Gook Moon  and Yuji Matsuda  and Takasada Shibauchi },
title = {Majorana-fermion origin of the planar thermal Hall effect in the {Kitaev} magnet {$\alpha$-RuCl$_3$}},
journal = {Science Advances},
volume = {10},
number = {11},
pages = {eadk3539},
year = {2024},
doi = {10.1126/sciadv.adk3539},
URL = {https://www.science.org/doi/abs/10.1126/sciadv.adk3539},}

@article{RuCl3_ultraclean_single_crystal,
  title={Magnetothermal transport in ultraclean single crystals of {Kitaev} magnet {$\alpha$-RuCl$_3$}},
  author={Xing, Y and Namba, R and Imamura, K and Ishihara, K and Suetsugu, S and Asaba, T and Hashimoto, K and Shibauchi, T and Matsuda, Y and Kasahara, Y},
  journal={npj Quantum Materials},
  volume={10},
  number={1},
  pages={33},
  year={2025},
  doi={10.1038/s41535-025-00749-4},
  publisher={Nature Publishing Group UK London}
}

@article{RuCl3_origin_Takagi,
    author = {Bruin, J. A. N. and Claus, R. R. and Matsumoto, Y. and Nuss, J. and Laha, S. and Lotsch, B. V. and Kurita, N. and Tanaka, H. and Takagi, H.},
    title = {Origin of oscillatory structures in the magnetothermal conductivity of the putative {Kitaev} magnet {$\alpha$-RuCl$_3$}},
    journal = {APL Materials},
    volume = {10},
    number = {9},
    pages = {090703},
    year = {2022},
    month = {09},
    issn = {2166-532X},
    doi = {10.1063/5.0101377},
    url = {https://doi.org/10.1063/5.0101377},
}

@article{RuCl3_stacking_disorder,
  title = {Stacking disorder and thermal transport properties of {$\alpha$-RuCl$_3$}},
  author = {Zhang, Heda and McGuire, Michael A. and May, Andrew F. and Chao, Hsin-Yun and Zheng, Qiang and Chi, Miaofang and Sales, Brian C. and Mandrus, David G. and Nagler, Stephen E. and Miao, Hu and Ye, Feng and Yan, Jiaqiang},
  journal = {Phys. Rev. Mater.},
  volume = {8},
  issue = {1},
  pages = {014402},
  numpages = {8},
  year = {2024},
  month = {Jan},
  publisher = {American Physical Society},
  doi = {10.1103/PhysRevMaterials.8.014402},
  url = {https://link.aps.org/doi/10.1103/PhysRevMaterials.8.014402}
}

@article{RuCl3_top_phase_diagram_Li_Ern_Chern,
  title = {Topological phase diagrams of in-plane field polarized {Kitaev} magnets},
  author = {Chern, Li Ern and Castelnovo, Claudio},
  journal = {Phys. Rev. B},
  volume = {109},
  issue = {18},
  pages = {L180407},
  numpages = {7},
  year = {2024},
  month = {May},
  publisher = {American Physical Society},
  doi = {10.1103/PhysRevB.109.L180407},
  url = {https://link.aps.org/doi/10.1103/PhysRevB.109.L180407}
}

@article{generalized_Kitaev_model_Heqiu,
  title = {Magnetic field induced topological transitions and thermal conductivity in a generalized {Kitaev} model},
  author = {Li, Heqiu and Kim, Yong Baek and Kee, Hae-Young},
  journal = {Phys. Rev. B},
  volume = {105},
  issue = {24},
  pages = {245142},
  numpages = {8},
  year = {2022},
  month = {Jun},
  publisher = {American Physical Society},
  doi = {10.1103/PhysRevB.105.245142},
  url = {https://link.aps.org/doi/10.1103/PhysRevB.105.245142}
}

@article{Mott_insulator_Kitaev_Jackeli,
  title = {Mott Insulators in the Strong Spin-Orbit Coupling Limit: From {Heisenberg} to a Quantum Compass and {Kitaev} Models},
  author = {Jackeli, G. and Khaliullin, G.},
  journal = {Phys. Rev. Lett.},
  volume = {102},
  issue = {1},
  pages = {017205},
  numpages = {4},
  year = {2009},
  month = {Jan},
  publisher = {American Physical Society},
  doi = {10.1103/PhysRevLett.102.017205},
  url = {https://link.aps.org/doi/10.1103/PhysRevLett.102.017205}
}

@article{Slack_1973_thermal_review,
title = {Nonmetallic crystals with high thermal conductivity},
journal = {Journal of Physics and Chemistry of Solids},
volume = {34},
number = {2},
pages = {321-335},
year = {1973},
issn = {0022-3697},
doi = {https://doi.org/10.1016/0022-3697(73)90092-9},
url = {https://www.sciencedirect.com/science/article/pii/0022369773900929},
author = {G.A. Slack}
}

@article{VanVleck_1940,
  title = {Paramagnetic Relaxation Times for Titanium and Chrome Alum},
  author = {Van Vleck, J. H.},
  journal = {Phys. Rev.},
  volume = {57},
  issue = {5},
  pages = {426--447},
  numpages = {0},
  year = {1940},
  month = {Mar},
  publisher = {American Physical Society},
  doi = {10.1103/PhysRev.57.426},
  url = {https://link.aps.org/doi/10.1103/PhysRev.57.426}
}

@article{spin_phonon_Mattuck_1960,
  title = {Spin-Phonon Interaction in Paramagnetic Crystals},
  author = {Mattuck, R. D. and Strandberg, M. W. P.},
  journal = {Phys. Rev.},
  volume = {119},
  issue = {4},
  pages = {1204--1217},
  numpages = {0},
  year = {1960},
  month = {Aug},
  publisher = {American Physical Society},
  doi = {10.1103/PhysRev.119.1204},
  url = {https://link.aps.org/doi/10.1103/PhysRev.119.1204}
}

@article{li2025thermal_review_2,
title = {Thermal transport in magnetic materials: A review},
journal = {Materials Today Electronics},
volume = {12},
pages = {100156},
year = {2025},
issn = {2772-9494},
doi = {https://doi.org/10.1016/j.mtelec.2025.100156},
url = {https://www.sciencedirect.com/science/article/pii/S2772949425000221},
author = {Shuchen Li and Shucheng Guo and Thomas Hoke and Xi Chen}
}

@article{thermal_conductivity_spin_chain_2003,
  title = {Thermal conductivity of spin-$\frac{1}{2}$ chains},
  author = {Shimshoni, E. and Andrei, N. and Rosch, A.},
  journal = {Phys. Rev. B},
  volume = {68},
  issue = {10},
  pages = {104401},
  numpages = {15},
  year = {2003},
  month = {Sep},
  publisher = {American Physical Society},
  doi = {10.1103/PhysRevB.68.104401},
  url = {https://link.aps.org/doi/10.1103/PhysRevB.68.104401}
}

@article{pseudoscalar_U1SL_RuCl3,
  title = {Pseudoscalar U(1) spin liquids in {$\alpha$-RuCl$_3$}},
  author = {Villadiego, Inti Sodemann},
  journal = {Phys. Rev. B},
  volume = {104},
  issue = {19},
  pages = {195149},
  numpages = {14},
  year = {2021},
  month = {Nov},
  publisher = {American Physical Society},
  doi = {10.1103/PhysRevB.104.195149},
  url = {https://link.aps.org/doi/10.1103/PhysRevB.104.195149}
}

@article{thermal_conductivity_spinon_PRB_2000,
  title = {Thermal conductivity and specific heat of the linear chain cuprate {${\mathrm{Sr}}_{2}{\mathrm{CuO}}_{3}$}: Evidence for thermal transport via spinons},
  author = {Sologubenko, A. V. and Felder, E. and Giann\`o, K. and Ott, H. R. and Vietkine, A. and Revcolevschi, A.},
  journal = {Phys. Rev. B},
  volume = {62},
  issue = {10},
  pages = {R6108(R)--R6111(R)},
  numpages = {0},
  year = {2000},
  month = {Sep},
  publisher = {American Physical Society},
  doi = {10.1103/PhysRevB.62.R6108},
  url = {https://link.aps.org/doi/10.1103/PhysRevB.62.R6108}
}

@article{thermal_conductivity_spin_ladder_PRL2000,
  title = {Thermal Conductivity of the Hole-Doped Spin Ladder System {${\mathrm{Sr}}_{14\ensuremath{-}\mathit{x}}{\mathrm{Ca}}_{\mathit{x}}{\mathrm{Cu}}_{24}\mathrm{O}_{41}$}},
  author = {Sologubenko, A. V. and Giann\'o, K. and Ott, H. R. and Ammerahl, U. and Revcolevschi, A.},
  journal = {Phys. Rev. Lett.},
  volume = {84},
  issue = {12},
  pages = {2714--2717},
  numpages = {0},
  year = {2000},
  month = {Mar},
  publisher = {American Physical Society},
  doi = {10.1103/PhysRevLett.84.2714},
  url = {https://link.aps.org/doi/10.1103/PhysRevLett.84.2714}
}

@article{magnon_heat_LaCuO_PRL2003,
  title = {Magnon Heat Transport in Doped {${\mathrm{L}\mathrm{a}}_{2}{\mathrm{C}\mathrm{u}\mathrm{O}}_{4}$}},
  author = {Hess, C. and B\"uchner, B. and Ammerahl, U. and Colonescu, L. and Heidrich-Meisner, F. and Brenig, W. and Revcolevschi, A.},
  journal = {Phys. Rev. Lett.},
  volume = {90},
  issue = {19},
  pages = {197002},
  numpages = {4},
  year = {2003},
  month = {May},
  publisher = {American Physical Society},
  doi = {10.1103/PhysRevLett.90.197002},
  url = {https://link.aps.org/doi/10.1103/PhysRevLett.90.197002}
}

@article{thermal_conductivity_spin_peierls_PRB1998,
  title = {Thermal conductivity of the spin-Peierls compound {${\mathrm{CuGeO}}_{3}$}},
  author = {Ando, Yoichi and Takeya, J. and Sisson, D. L. and Doettinger, S. G. and Tanaka, I. and Feigelson, R. S. and Kapitulnik, A.},
  journal = {Phys. Rev. B},
  volume = {58},
  issue = {6},
  pages = {R2913--R2916},
  numpages = {0},
  year = {1998},
  month = {Aug},
  publisher = {American Physical Society},
  doi = {10.1103/PhysRevB.58.R2913},
  url = {https://link.aps.org/doi/10.1103/PhysRevB.58.R2913}
}

@article{spinon_transport_PbCuTe2O6_PRL2023,
  title = {Spinon Heat Transport in the Three-Dimensional Quantum Magnet {${\mathrm{PbCuTe}}_{2}{\mathrm{O}}_{6}$}},
  author = {Hong, Xiaochen and Gillig, Matthias and Hanna, Abanoub R. N. and Chillal, Shravani and Islam, A. T. M. Nazmul and Lake, Bella and B\"uchner, Bernd and Hess, Christian},
  journal = {Phys. Rev. Lett.},
  volume = {131},
  issue = {25},
  pages = {256701},
  numpages = {6},
  year = {2023},
  month = {Dec},
  publisher = {American Physical Society},
  doi = {10.1103/PhysRevLett.131.256701},
  url = {https://link.aps.org/doi/10.1103/PhysRevLett.131.256701}
}

@article{heat_current_spin_liquid_Science_2010,
author = {Minoru Yamashita  and Norihito Nakata  and Yoshinori Senshu  and Masaki Nagata  and Hiroshi M. Yamamoto  and Reizo Kato  and Takasada Shibauchi  and Yuji Matsuda },
title = {Highly Mobile Gapless Excitations in a Two-Dimensional Candidate Quantum Spin Liquid},
journal = {Science},
volume = {328},
number = {5983},
pages = {1246-1248},
year = {2010},
doi = {10.1126/science.1188200},
URL = {https://www.science.org/doi/abs/10.1126/science.1188200}
}

@article{qsl_heat_1TTaS2_PRR_2020,
  title = {Effect of quenched disorder on the quantum spin liquid state of the triangular-lattice antiferromagnet {$1T\ensuremath{-}{\mathrm{TaS}}_{2}$}},
  author = {Murayama, H. and Sato, Y. and Taniguchi, T. and Kurihara, R. and Xing, X. Z. and Huang, W. and Kasahara, S. and Kasahara, Y. and Kimchi, I. and Yoshida, M. and Iwasa, Y. and Mizukami, Y. and Shibauchi, T. and Konczykowski, M. and Matsuda, Y.},
  journal = {Phys. Rev. Res.},
  volume = {2},
  issue = {1},
  pages = {013099},
  numpages = {9},
  year = {2020},
  month = {Jan},
  publisher = {American Physical Society},
  doi = {10.1103/PhysRevResearch.2.013099},
  url = {https://link.aps.org/doi/10.1103/PhysRevResearch.2.013099}
}

@article{thermal_Hall_neutral_spin_Science_2015,
author = {Max Hirschberger  and Jason W. Krizan  and R. J. Cava  and N. P. Ong },
title = {Large thermal Hall conductivity of neutral spin excitations in a frustrated quantum magnet},
journal = {Science},
volume = {348},
number = {6230},
pages = {106-109},
year = {2015},
doi = {10.1126/science.1257340},
URL = {https://www.science.org/doi/abs/10.1126/science.1257340}}

@article{heat_qsl_kagome_volborthite_pnas2016,
author = {Daiki Watanabe  and Kaori Sugii  and Masaaki Shimozawa  and Yoshitaka Suzuki  and Takeshi Yajima  and Hajime Ishikawa  and Zenji Hiroi  and Takasada Shibauchi  and Yuji Matsuda  and Minoru Yamashita },
title = {Emergence of nontrivial magnetic excitations in a spin-liquid state of kagomé volborthite},
journal = {Proceedings of the National Academy of Sciences},
volume = {113},
number = {31},
pages = {8653-8657},
year = {2016},
doi = {10.1073/pnas.1524076113},
URL = {https://www.pnas.org/doi/abs/10.1073/pnas.1524076113}
}

@article{spin_thermal_Hall_Kogome_AFM_2018,
  title = {Spin Thermal Hall Conductivity of a Kagome Antiferromagnet},
  author = {Doki, Hayato and Akazawa, Masatoshi and Lee, Hyun-Yong and Han, Jung Hoon and Sugii, Kaori and Shimozawa, Masaaki and Kawashima, Naoki and Oda, Migaku and Yoshida, Hiroyuki and Yamashita, Minoru},
  journal = {Phys. Rev. Lett.},
  volume = {121},
  issue = {9},
  pages = {097203},
  numpages = {6},
  year = {2018},
  month = {Aug},
  publisher = {American Physical Society},
  doi = {10.1103/PhysRevLett.121.097203},
  url = {https://link.aps.org/doi/10.1103/PhysRevLett.121.097203}
}

@article{thermall_hall_cuprate_Mott_insulator_NC2020,
  title={Thermal Hall conductivity in the cuprate Mott insulators {Nd$_2$CuO$_4$} and {Sr$_2$CuO$_2$Cl$_2$}},
  author={Boulanger, Marie-Eve and Grissonnanche, Ga{\"e}l and Badoux, Sven and Allaire, Andr{\'e}anne and Lefran{\c{c}}ois, {\'E}tienne and Legros, Ana{\"e}lle and Gourgout, Adrien and Dion, Maxime and Wang, CH and Chen, XH and others},
  journal={Nature communications},
  volume={11},
  number={1},
  pages={5325},
  year={2020},
  publisher={Nature Publishing Group UK London},
  doi = {10.1038/s41467-020-18881-z},
  url={https://doi.org/10.1038/s41467-020-18881-z}
}

@article{thermal_conductivity_qsl_Nature_2009,
  title={Thermal-transport measurements in a quantum spin-liquid state of the frustrated triangular magnet {$(\text {BEDT-TTF})_2 \mathrm{Cu}_2(\mathrm{CN})_3$}},
  author={Yamashita, Minoru and Nakata, Norihito and Kasahara, Yuichi and Sasaki, Takahiko and Yoneyama, Naoki and Kobayashi, Norio and Fujimoto, Satoshi and Shibauchi, Takasada and Matsuda, Yuji},
  journal={Nature Physics},
  volume={5},
  number={1},
  pages={44--47},
  year={2009},
  publisher={Nature Publishing Group UK London},
  doi={10.1038/nphys1134},
  url = {https://doi.org/10.1038/nphys1134}
}

@article{qsl_thermal_conductivity_PRB2025,
  title = {Possible quantum spin liquid state of {${\mathrm{CeTa}}_{7}{\mathrm{O}}_{19}$}},
  author = {Li, N. and Rutherford, A. and Wang, Y. Y. and Liang, H. and Zhou, Y. and Sun, Y. and Wu, D. D. and Chen, P. F. and Li, Q. J. and Wang, H. and Xie, W. and Choi, E. S. and Zhang, S. Z. and Lee, M. and Zhou, H. D. and Sun, X. F.},
  journal = {Phys. Rev. B},
  volume = {111},
  issue = {9},
  pages = {094414},
  numpages = {8},
  year = {2025},
  month = {Mar},
  publisher = {American Physical Society},
  doi = {10.1103/PhysRevB.111.094414},
  url = {https://link.aps.org/doi/10.1103/PhysRevB.111.094414}
}

@article{Ising-qsl_PRB2024,
  title = {{Ising}-type quantum spin liquid state in {${\mathrm{PrMgAl}}_{11}{\mathrm{O}}_{19}$}},
  author = {Li, N. and Rutherford, A. and Wang, Y. Y. and Liang, H. and Li, Q. J. and Zhang, Z. J. and Wang, H. and Xie, W. and Zhou, H. D. and Sun, X. F.},
  journal = {Phys. Rev. B},
  volume = {110},
  issue = {13},
  pages = {134401},
  numpages = {9},
  year = {2024},
  month = {Oct},
  publisher = {American Physical Society},
  doi = {10.1103/PhysRevB.110.134401},
  url = {https://link.aps.org/doi/10.1103/PhysRevB.110.134401}
}

@article{triangular_qsl_thermal_PRR2024,
  title = {Gapped quantum spin liquid in a triangular-lattice {Ising}-type antiferromagnet {${\mathrm{PrMgAl}}_{11}{\mathrm{O}}_{19}$}},
  author = {Tu, C. P. and Ma, Z. and Wang, H. R. and Jiao, Y. H. and Dai, D. Z. and Li, S. Y.},
  journal = {Phys. Rev. Res.},
  volume = {6},
  issue = {4},
  pages = {043147},
  numpages = {9},
  year = {2024},
  month = {Nov},
  publisher = {American Physical Society},
  doi = {10.1103/PhysRevResearch.6.043147},
  url = {https://link.aps.org/doi/10.1103/PhysRevResearch.6.043147}
}

@article{review_spin_chain_heat_2021,
  title = {Finite-temperature transport in one-dimensional quantum lattice models},
  author = {Bertini, B. and Heidrich-Meisner, F. and Karrasch, C. and Prosen, T. and Steinigeweg, R. and \ifmmode \check{Z}\else \v{Z}\fi{}nidari\ifmmode \check{c}\else \v{c}\fi{}, M.},
  journal = {Rev. Mod. Phys.},
  volume = {93},
  issue = {2},
  pages = {025003},
  numpages = {71},
  year = {2021},
  month = {May},
  publisher = {American Physical Society},
  doi = {10.1103/RevModPhys.93.025003},
  url = {https://link.aps.org/doi/10.1103/RevModPhys.93.025003}
}

@article{magnetoelastic_coupling_RuCl3_2021,
  title = {Magnetoelastic coupling and effects of uniaxial strain in {$\ensuremath{\alpha}\ensuremath{-}{\mathrm{RuCl}}_{3}$} from first principles},
  author = {Kaib, David A. S. and Biswas, Sananda and Riedl, Kira and Winter, Stephen M. and Valent\'{\i}, Roser},
  journal = {Phys. Rev. B},
  volume = {103},
  issue = {14},
  pages = {L140402},
  numpages = {6},
  year = {2021},
  month = {Apr},
  publisher = {American Physical Society},
  doi = {10.1103/PhysRevB.103.L140402},
  url = {https://link.aps.org/doi/10.1103/PhysRevB.103.L140402}
}

@article{phase_diagram_XXZ_chain,
  title = {Spin-$\frac{1}{2}$ {XXZ} {Heisenberg} chain in a longitudinal magnetic field},
  author = {Rakov, Mykhailo V. and Weyrauch, Michael},
  journal = {Phys. Rev. B},
  volume = {100},
  issue = {13},
  pages = {134434},
  numpages = {9},
  year = {2019},
  month = {Oct},
  publisher = {American Physical Society},
  doi = {10.1103/PhysRevB.100.134434},
  url = {https://link.aps.org/doi/10.1103/PhysRevB.100.134434}
}

@article{heat_transport_low_quantum_magnet_review_Hess_2019,
title = {Heat transport of cuprate-based low-dimensional quantum magnets with strong exchange coupling},
journal = {Physics Reports},
volume = {811},
pages = {1-38},
year = {2019},
issn = {0370-1573},
doi = {https://doi.org/10.1016/j.physrep.2019.02.004},
url = {https://www.sciencedirect.com/science/article/pii/S0370157319301036},
author = {Christian Hess}
}

@article{large_magnon_heat_Cu_2017,
  title = {Ballistic magnon heat conduction and possible Poiseuille flow in the helimagnetic insulator {${\mathrm{Cu}}_{2}{\mathrm{OSeO}}_{3}$}},
  author = {Prasai, N. and Trump, B. A. and Marcus, G. G. and Akopyan, A. and Huang, S. X. and McQueen, T. M. and Cohn, J. L.},
  journal = {Phys. Rev. B},
  volume = {95},
  issue = {22},
  pages = {224407},
  numpages = {9},
  year = {2017},
  month = {Jun},
  publisher = {American Physical Society},
  doi = {10.1103/PhysRevB.95.224407},
  url = {https://link.aps.org/doi/10.1103/PhysRevB.95.224407}
}

@article{thermal_transport_probing_quantum_material_review_Li_2020, title={Thermal transport for probing quantum materials}, volume={45}, DOI={10.1557/mrs.2020.124}, number={5}, journal={MRS Bulletin}, author={Li, Mingda and Chen, Gang}, year={2020}, pages={348–356},url={https://doi.org/10.1557/mrs.2020.124}}

@article{thermal_Hall_review_2024,
title = {Thermal Hall effects in quantum magnets},
journal = {Physics Reports},
volume = {1070},
pages = {1-59},
year = {2024},
issn = {0370-1573},
doi = {https://doi.org/10.1016/j.physrep.2024.03.004},
url = {https://www.sciencedirect.com/science/article/pii/S0370157324001194},
author = {Xiao-Tian Zhang and Yong Hao Gao and Gang Chen}
}

@article{
QSL_review_2020,
author = {C. Broholm  and R. J. Cava  and S. A. Kivelson  and D. G. Nocera  and M. R. Norman  and T. Senthil },
title = {Quantum spin liquids},
journal = {Science},
volume = {367},
number = {6475},
pages = {eaay0668},
year = {2020},
doi = {10.1126/science.aay0668},
URL = {https://www.science.org/doi/abs/10.1126/science.aay0668}}

@article{YIG_magnons_field_PRB_2015,
  title = {Thermal properties of magnons in yttrium iron garnet at elevated magnetic fields},
  author = {Rezende, S. M. and L\'opez Ortiz, J. C.},
  journal = {Phys. Rev. B},
  volume = {91},
  issue = {10},
  pages = {104416},
  numpages = {6},
  year = {2015},
  month = {Mar},
  publisher = {American Physical Society},
  doi = {10.1103/PhysRevB.91.104416},
  url = {https://link.aps.org/doi/10.1103/PhysRevB.91.104416}
}

@article{YIG_large_field_PRB_2020,
  title = {Thermal transport in yttrium iron garnet at very high magnetic fields},
  author = {Ratkovski, D. R. and Balicas, L. and Bangura, A. and Machado, F. L. A. and Rezende, S. M.},
  journal = {Phys. Rev. B},
  volume = {101},
  issue = {17},
  pages = {174442},
  numpages = {6},
  year = {2020},
  month = {May},
  publisher = {American Physical Society},
  doi = {10.1103/PhysRevB.101.174442},
  url = {https://link.aps.org/doi/10.1103/PhysRevB.101.174442}
}

@article{magnon_mean_free_path_YIG_PRB_2014,
  title = {Magnon thermal mean free path in yttrium iron garnet},
  author = {Boona, Stephen R. and Heremans, Joseph P.},
  journal = {Phys. Rev. B},
  volume = {90},
  issue = {6},
  pages = {064421},
  numpages = {8},
  year = {2014},
  month = {Aug},
  publisher = {American Physical Society},
  doi = {10.1103/PhysRevB.90.064421},
  url = {https://link.aps.org/doi/10.1103/PhysRevB.90.064421}
}

@article{thermal_CrCl3_PRR_2020,
  title = {Giant thermal magnetoconductivity in ${\mathrm{CrCl}}_{3}$ and a general model for spin-phonon scattering},
  author = {Pocs, Christopher A. and Leahy, Ian A. and Zheng, Hao and Cao, Gang and Choi, Eun-Sang and Do, S.-H. and Choi, Kwang-Yong and Normand, B. and Lee, Minhyea},
  journal = {Phys. Rev. Res.},
  volume = {2},
  issue = {1},
  pages = {013059},
  numpages = {13},
  year = {2020},
  month = {Jan},
  publisher = {American Physical Society},
  doi = {10.1103/PhysRevResearch.2.013059},
  url = {https://link.aps.org/doi/10.1103/PhysRevResearch.2.013059}
}

@article{resonant_phonon_scattering_Ni3TeO6_Heejun_PRB2022,
  title = {Diagonal and off-diagonal thermal conduction with resonant phonon scattering in {${\mathrm{Ni}}_{3}\mathrm{Te}{\mathrm{O}}_{6}$}},
  author = {Yang, Heejun and Xu, Xianghan and Lee, Jun Han and Oh, Yoon Seok and Cheong, Sang-Wook and Park, Je-Geun},
  journal = {Phys. Rev. B},
  volume = {106},
  issue = {14},
  pages = {144417},
  numpages = {9},
  year = {2022},
  month = {Oct},
  publisher = {American Physical Society},
  doi = {10.1103/PhysRevB.106.144417},
  url = {https://link.aps.org/doi/10.1103/PhysRevB.106.144417}
}

@article{magnon_phonon_Walton_PRB1977,
  title = {Effect of magnon-phonon thermal relaxation on heat transport by magnons},
  author = {Sanders, D. J. and Walton, D.},
  journal = {Phys. Rev. B},
  volume = {15},
  issue = {3},
  pages = {1489--1494},
  numpages = {0},
  year = {1977},
  month = {Feb},
  publisher = {American Physical Society},
  doi = {10.1103/PhysRevB.15.1489},
  url = {https://link.aps.org/doi/10.1103/PhysRevB.15.1489}
}

@article{spin_phonon_thermal_Walton_PRB_1970,
  title = {Effect of the Spin-Phonon Interaction on the Thermal Conductivity},
  author = {Walton, D.},
  journal = {Phys. Rev. B},
  volume = {1},
  issue = {3},
  pages = {1234--1242},
  numpages = {0},
  year = {1970},
  month = {Feb},
  publisher = {American Physical Society},
  doi = {10.1103/PhysRevB.1.1234},
  url = {https://link.aps.org/doi/10.1103/PhysRevB.1.1234}
}

@article{giant_supression_phonon_heat_BiCu2PO6_Sci_reports_2016,
  title={Giant suppression of phononic heat transport in a quantum magnet {BiCu$_2$PO$_6$}},
  author={Jeon, Byung-Gu and Koteswararao, B and Park, CB and Shu, GJ and Riggs, SC and Moon, Eun-Gook and Chung, SB and Chou, FC and Kim, Kee Hoon},
  journal={Scientific reports},
  volume={6},
  number={1},
  pages={36970},
  year={2016},
  publisher={Nature Publishing Group UK London},
  doi={10.1038/srep36970},
  url={https://doi.org/10.1038/srep36970}
}

@article{Churchill_2024_prl,
  title = {Transforming from {Kitaev} to Disguised {Ising} Chain: Application to {${\mathrm{CoNb}}_{2}{\mathrm{O}}_{6}$}},
  author = {Churchill, Derek and Kee, Hae-Young},
  journal = {Phys. Rev. Lett.},
  volume = {133},
  issue = {5},
  pages = {056703},
  numpages = {6},
  year = {2024},
  month = {Jul},
  publisher = {American Physical Society},
  doi = {10.1103/PhysRevLett.133.056703},
  url = {https://link.aps.org/doi/10.1103/PhysRevLett.133.056703}
}

@article{XXZ_alter_PRB_2024,
  title = {Phase diagram and topology of the {XXZ} chain with alternating bonds and staggered magnetic field},
  author = {M\'arquez, B. F. and Aucar Boidi, N. and Hallberg, K. and Aligia, A. A.},
  journal = {Phys. Rev. B},
  volume = {109},
  issue = {23},
  pages = {235143},
  numpages = {8},
  year = {2024},
  month = {Jun},
  publisher = {American Physical Society},
  doi = {10.1103/PhysRevB.109.235143},
  url = {https://link.aps.org/doi/10.1103/PhysRevB.109.235143}
}

@article{MengxingYe_PHV_magnetic_insulators,
  title={Phonon Hall viscosity in magnetic insulators},
  author={Ye, Mengxing and Savary, Lucile and Balents, Leon},
  journal={arXiv preprint},
  year={2021},
  doi={10.48550/arXiv.2103.04223}
}

@article{fractionalized_ultrasound_PRB_2024,
  title = {Fractionalized excitations probed by ultrasound},
  author = {Hauspurg, A. and Zherlitsyn, S. and Helm, T. and Felea, V. and Wosnitza, J. and Tsurkan, V. and Choi, K.-Y. and Do, S.-H. and Ye, Mengxing and Brenig, Wolfram and Perkins, Natalia B.},
  journal = {Phys. Rev. B},
  volume = {109},
  issue = {14},
  pages = {144415},
  numpages = {11},
  year = {2024},
  month = {Apr},
  publisher = {American Physical Society},
  doi = {10.1103/PhysRevB.109.144415},
  url = {https://link.aps.org/doi/10.1103/PhysRevB.109.144415}
}

@article{phonon_dynamics_KSL_mengxing_ye,
  title = {Phonon dynamics in the {Kitaev} spin liquid},
  author = {Ye, Mengxing and Fernandes, Rafael M. and Perkins, Natalia B.},
  journal = {Phys. Rev. Res.},
  volume = {2},
  issue = {3},
  pages = {033180},
  numpages = {14},
  year = {2020},
  month = {Aug},
  publisher = {American Physical Society},
  doi = {10.1103/PhysRevResearch.2.033180},
  url = {https://link.aps.org/doi/10.1103/PhysRevResearch.2.033180}
}

@article{heat_hybridization_Minhyea_Lee_PRR_2025,
  title = {Heat conduction in magnetic insulators via hybridization of acoustic phonons and spin-flip excitations},
  author = {Pocs, Christopher A. and Leahy, Ian A. and Xing, Jie and Choi, Eun Sang and Sefat, Athena S. and Hermele, Michael and Lee, Minhyea},
  journal = {Phys. Rev. Res.},
  volume = {7},
  issue = {2},
  pages = {L022007},
  numpages = {7},
  year = {2025},
  month = {Apr},
  publisher = {American Physical Society},
  doi = {10.1103/PhysRevResearch.7.L022007},
  url = {https://link.aps.org/doi/10.1103/PhysRevResearch.7.L022007}
}

@article{comprehensive_KJGamma_chain_2020_Affleck,
  title = {Comprehensive study of the phase diagram of the spin-$\frac{1}{2}$ {Kitaev}-{Heisenberg}-{Gamma} chain},
  author = {Yang, Wang and Nocera, Alberto and Affleck, Ian},
  journal = {Phys. Rev. Res.},
  volume = {2},
  issue = {3},
  pages = {033268},
  numpages = {24},
  year = {2020},
  month = {Aug},
  publisher = {American Physical Society},
  doi = {10.1103/PhysRevResearch.2.033268},
  url = {https://link.aps.org/doi/10.1103/PhysRevResearch.2.033268}
}

@article{Cen2025PRB,
  title = {Intermediate phases in {$\ensuremath{\alpha}\text{\ensuremath{-}}{\mathrm{RuCl}}_{3}$} under in-plane magnetic field via interlayer spin interactions},
  author = {Cen, Jiefu and Kee, Hae-Young},
  journal = {Phys. Rev. B},
  volume = {112},
  issue = {2},
  pages = {024419},
  numpages = {12},
  year = {2025},
  month = {Jul},
  publisher = {American Physical Society},
  doi = {10.1103/tr8c-pxbr},
  url = {https://link.aps.org/doi/10.1103/tr8c-pxbr}
}

\end{document}